\documentclass[useAMS]{mn2e}
\usepackage{graphicx}
\usepackage{latexsym}
\usepackage {keyval}
\usepackage{amsmath}
\title[``Head/tail'' plasmon model]
{A ``head/tail'' plasmon model with a Hubble law velocity profile}
\author[Raga, A. C., Rodr\'\i guez-Gonz\'alez, A.,
  Hern\'andez-Mart\'\i nez, L.,
  Cant\'o, J., Castellanos-Ram\'\i rez, A.]{
  A. C. Raga$^{1,2}$\thanks{E-mail: raga@nucleares.unam.mx},
  A. Rodr\'\i guez-Gonz\'alez$^1$,
  L. Hern\'andez-Mart\'\i nez$^1$, \and
  J. Cant\'o$^3$, A. Castellanos-Ram\'\i rez$^3$\\
$^1$Instituto de Ciencias Nucleares,
Universidad Nacional Aut\'onoma de M\'exico,
Ap. 70-543, 04510 D. F., M\'exico\\
$^2$ Instituto de Investigaci\'on en Ciencias F\'\i sicas
y Matem\'aticas,
USAC, Ciudad Universitaria, Zona 12, Guatemala\\
$^3$Instituto de Astronom\'\i a, Universidad Nacional Aut\'onoma de M\'exico,
Ap. 70-468, 04510 D. F., M\'exico\\
}
\begin{document}

\date{}

\pagerange{\pageref{firstpage}--\pageref{lastpage}} \pubyear{2019}

\maketitle

\label{firstpage}

\begin{abstract}
  We present a model of a hypersonic, collimated, ``single pulse'' outflow,
  produced by an event with an ejection velocity
  that first grows, reaches a peak, and then decreases again to zero
  velocity in a finite time (simultaneously, the ejection density
  can have an arbitrary time-variability). We obtain
  a flow with a leading ``head'' and a trailing ``tail'
  that for times greater than the width of the pulse
  develops a linear, ``Hubble law'' velocity vs. position.
  We present an analytic model for
  a simple pulse with a parabolic ejection velocity vs. time and
  time-independent mass-loss rate, and compare it to
  an axisymmetric gasdynamic simulation with parameters appropriate
  for fast knots in planetary nebulae. This ``head/tail plasmon''
  flow might be applicable to other high-velocity clumps with ``Hubble law''
  tails.
\end{abstract}


\section{Introduction}

A pattern that is sometimes seen in collimated stellar outflows
is a high-velocity, compact ``clump'', joined to the outflow
source by fainter emission with a linear ramp of increasing
velocity as a function of distance from the source. This
results in striking ``position-velocity'' (PV) diagrams (obtained,
e.g., from long-slit, high resolution spectra or from millimetre
interferometric ``position-velocity cubes'') with a linear
ramp ending in a bright, high-velocity condensation.

Alcolea et al. (2001) proposed that clumps with ``Hubble law tails''
(observed in the CO emission of a collimated, protoplanetary nebula
outflow) are produced in
``explosive events'' (i.e., with a duration much shorter than the
evolutionary time of the outflow). A ``velocity sorting''
mechanism (with higher velocity material racing ahead of slower
ejecta) would then produce the observed linear velocity vs. position
structure of the tails.

The most dramatic example of ``Hubble law tail clumps'' is of
course found in the molecular fingers pointing away from the
Orion BN-KL region (see, e.g., Allen \& Burton 1993; Zapata et al. 2011;
Bally et al. 2017). The $\sim 100$ fingers
all show CO emission with linearly increasing radial velocities away
from the outflow centre, and terminate in compact clumps (observed
in H$_2$ and in optical atomic/ionic lines).

Dennis et al. (2008) presented numerical simulations of variable jets
and of outflows composed of discrete ``clumps'', and conclude
that the clump-like outflows produce a compact ``head'' (i.e.,
the clump), followed by a tail of decreasing velocity material.
They favour this ``clump scenario'' for explaining the
observed ``Hubble law'' PV diagrams of clumps in planetary nebulae (PNe).
However, even though they obtain trails of decreasing velocity
material (between the clumps and the outflow source), these
trails do not show either the length or the very dramatic
linear velocity vs. position signatures of the observed clumps.

In the present paper we explore a scenario similar to the one of
Dennis et al. (2008), but instead of imagining a ``clump''
ejected from the source (with a well defined ejection velocity), we
propose a ``single pulse''-type ejection velocity (and density)
variability. Basically, during a finite time the source ejects material
first at increasing velocities, then reaching a maximum ejection velocity,
and finally decreasing down to zero. In principle, within this
``ejection episode'', the density of the ejected material could
also vary in an arbitrary way,

In sections 2-5
we present a simple analytic model of the
resulting ``head/tail plasmon'' flow,
calculate its time-evolution and obtain predicted
PV diagrams. We also compute an
axisymmetric numerical simulation of this
flow (with parameters appropriate for a clump in a PN),
and compare it with our analytic model (section 6).

\section{The plasmon model}

\subsection{Centre of mass equation of motion}

Let us consider a cylindrical outflow, with an ejection ``pulse''
beginning at an ejection time $\tau=-\tau_0$ and ending at $\tau=\tau_0$.
This pulse has an arbitrary ejection density $\rho_0(\tau)$ and
an ejection velocity $u_0(\tau)=0$ for $|\tau|\geq\tau_0$
and $u_0(\tau)>0$ for $|\tau|<\tau_0$, This ejection
travels into a stationary environment of uniform density $\rho_a$.

Clearly, as the ejection pulse evolves, the faster material ejected
at later times catches up with the slower, earlier ejection, producing
a shock wave. Also, a second shock wave (i.e., the bow shock) is
produced in the interaction of the jet with the surrounding
environment. This working surface is the ``head'' of the plasmon.

At later times, the ``tail'' region between the ``head'' and the source
is filled with the material ejected in the tail of the ejection pulse.
and has a velocity that increases out to the position of the ``head''.
We call this flow (shown in the schematic diagram of Figure 1)
the ``head/tail plasmon'' in order to difference
it from the plasmon of De Young \& Axford (1967).

\begin{figure}
\centering
\includegraphics[width=5cm]{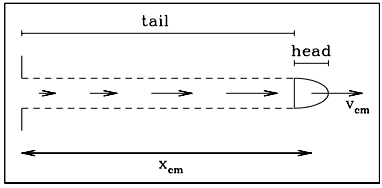}
\caption{Schematic diagram showing the ``head/tail'' plasmon. The head
  (at a distance $x_{cm}$ from the source)
  travels at a velocity $v_{cm}$ along the $x$-axis,
  and the tail of unshocked material eventually develops a velocity
  stratification with lower velocities closer to the outflow source.}
\label{fig1}
\end{figure}

Using the ``centre of mass'' formalism of Cant\'o et al. (2000),
we will assume that:
\begin{enumerate}
\item before reaching the working surface
  the ejected material is free-streaming (as appropriate
  for a hypersonic flow),
\item the working surface has a position that coincides with
  the centre of mass of the material within it (calculated
  as if the material were still free-streaming).
\end{enumerate}
This latter point is correct if the working surface can be
seen as an inelastic merger of flow parcels.

With these two points, the position of the head (i.e.,
the working surface) coincides with the centre of mass:
\begin{equation}
  x_{cm}=\frac{\int_{-\tau_0}^\tau \rho_0 x_j u_0 d\tau'+\int_0^{x_{cm}}\rho_a x dx}
  {\int_{-\tau_0}^\tau \rho_0u_0d\tau'+\int_0^{x_{cm}}\rho_adx}\,,
  \label{xcm}
\end{equation}
where $u_0(\tau')$ and $\rho_0(\tau')$ are the time-dependent
ejection velocity and density (respectively), and $\rho_a(x)$ is
the environmental density. The outflow source is assumed to be
at $x=0$, and the cylindrical ejection is parallel to the
$x$-axis (see Figure 1).

The position $x_j$ of the fluid parcels (if they had not merged)
is given by the free-streaming relation:
\begin{equation}
  x_j=(t-\tau')u_0(\tau')\,,
  \label{xj}
\end{equation}
where $t$ is the ``evolutionary time'' (different from the ejection
time $\tau'$, satisfying the condition $t\geq \tau'$). The
upper limit $\tau$ of the integrals is given by the free-streaming
flow condition:
\begin{equation}
  x_{cm}=(t-\tau)u_0(\tau)\,,
  \label{xttau}
\end{equation}
for the ejected fluid parcels currently (i.e., at time $t$) entering the
working surface.

Now, combining equations (\ref{xcm}-\ref{xttau}), and considering
a uniform environment (with $\rho_a=const.$), we obtain:
$$
  \frac{\rho_a x_{cm}^2}{2}+x_{cm}
  \,\left[\int_{-\tau_0}^\tau \rho_0u_0d\tau'-
    \frac{1}{u_0(\tau)} \int_{-\tau_0}^\tau \rho_0u_0^2 d\tau'\right]=
$$
\begin{equation} 
  \tau\int_{\tau_0}^\tau\rho_0 u_0^2d\tau'-\int_{\tau_0}^\tau \tau'\rho_0u_0^2d\tau'\,,
  \label{xcm1}
\end{equation}
which, once the appropriate integrals over $\tau'$ have been carried
out, is a quadratic equation which gives us $x_{cm}(\tau)$. If we
want to know the position of the working surface as a function
of the evolutionary time $t$, we can calculate $t$ as a function
of $\tau$ and $x_{cm}(\tau)$ from equation (\ref{xttau}).

\subsection{Solution for a parabolic $u_0(\tau)$ pulse with constant
  mass loss rate}

Let us now consider an ejection velocity pulse with
$u_0(\tau)=0$ for $|\tau|\geq\tau_0$ and:
\begin{equation}
  u_0(\tau)=v_0\left[1-\left(\frac{\tau}{\tau_0}\right)^2\right]\,;\,\,\,
    {\rm for}\,\,|\tau|<\tau_0\,,
    \label{u0q}
\end{equation}
a parabola that goes to zero at $\tau=\pm \tau_0$ and has
a peak velocity $v_0$ at $\tau=0$. For the ejection density
$\rho_0(\tau)$, we assume that it is proportional
to the inverse of the ejection velocity, so that the
mass loss rate (per unit area)
\begin{equation}
  {\dot m}=\rho_0(\tau)u_0(\tau)\,,
  \label{mdot}
\end{equation}
is time-independent. However, an arbitrary ejection density
variability could be considered within our analytic framework.

With $u_0(\tau)$ and $\rho_0(\tau)$ given by equations
(\ref{u0q}-\ref{mdot}) we compute the integrals in equation (\ref{xcm1}),
obtaining:
\begin{equation}
  \sigma\left(\frac{x_{cm}}{v_0\tau_0}\right)^2+
  f\left(\frac{\tau}{\tau_0}\right)\,\frac{x_{cm}}{v_0\tau_0}=
  g\left(\frac{\tau}{\tau_0}\right)\,,
  \label{xcm3}
\end{equation}
with
\begin{equation}
  \sigma\equiv \frac{\rho_av_0}{2{\dot m}}\,,
  \label{acm1}
\end{equation}
\begin{equation}
f(\eta)\equiv \frac{(2\eta-1)(\eta+1)}{3(\eta-1)}\,;\,\,\,\,\,
g(\eta)=\frac{(3-\eta)(\eta+1)^3}{12}\,.
  \label{acm3}
\end{equation}

\section{The ``free plasmon'', $\sigma=0$ case}

In the $\sigma\to 0$ limit
(see equation \ref{acm1}) of a very low density environment,
equation (\ref{xcm3}) has the solution:
\begin{equation}
  \frac{x_{cm}}{v_0\tau_0}=\frac{g(\tau/\tau_0)}{f(\tau/\tau_0)}\,,
  \label{xcm5}
\end{equation}
with $f$ and $g$ given by equation (\ref{acm3}).

Substituting equation (\ref{acm3}) in (\ref{xcm5}) we
obtain:
\begin{equation}
  \frac{x_{cm}}{v_0\tau_0}=\frac{(3-\eta)(\eta-1)(\eta+1)^2}
       {4(2\eta-1)}\,,
       \label{xcm55}
\end{equation}
where $\eta=\tau/\tau_0$.
Using the free-streaming flow condition (equation \ref{xttau}),
and equations (\ref{u0q}) and (\ref{xcm55}) we obtain:
\begin{equation}
  \frac{t}{\tau_0}=\frac{3(\eta-1)(1+3\eta)}{4(2\eta-1)}\,.
  \label{ttau6}
\end{equation}
Clearly, both $x_{cm}$ and $t$ $\to \infty$ for $\tau\to \tau_0/2$
(see equations \ref{xcm55}-\ref{ttau6}).

The velocity $v_{cm}=dx_{cm}/dt$ can be obtained from equations
(\ref{xcm55}-\ref{ttau6}):
\begin{equation}
  \frac{v_{cm}}{v_0}=\frac{1}{3}(2+\eta-\eta^2)\,.
    \label{vcm}
    \end{equation}
Therefore, for $t\to \infty$ ($\tau\to \tau_0/2$)
the plasmon head reaches an asymptotic velocity
\begin{equation}
  v_a=u_0\left(\frac{\tau_0}{2}\right)=\frac{3}{4}v_0\,.
  \label{va}
\end{equation}
This result implies that the material ejected in the part
of the pulse with $\tau>\tau_0/2$ (see equation \ref{u0q}) never
reaches the plasmon head.
In the ($t\to \infty$, $\tau\to \tau_0/2$) asymptotic regime,
the head of the plasmon has a mass (per unit area)
$m_{h,a}=3{\dot m}\tau_0/2$ and the tail has a mass
$m_{t,a}={\dot m}\tau_0/2$. Therefore, out of the total ejected mass
$m_{tot}=2{\dot m}\tau_0$, a fraction of 3/4 ends up in the head
and 1/4 in the tail.

In Figure 2 we plot $x_{cm}/(v_0\tau_0)$ (see equation \ref{xcm5})
as a function of $t$ (which is obtained from $\tau,\,x_{cm}(\tau)$
through equation \ref{xttau}). We also plot the velocity
$v_{cm}$ (given by equation \ref{vcm}) as a function of the
evolutionary time $t$. From this figure it is clear that the plasmon
head first accelerates, and for $t>\tau_0$ starts to approach
the asymptotic velocity given by equation (\ref{va}).

\begin{figure}
\centering
\includegraphics[width=6cm]{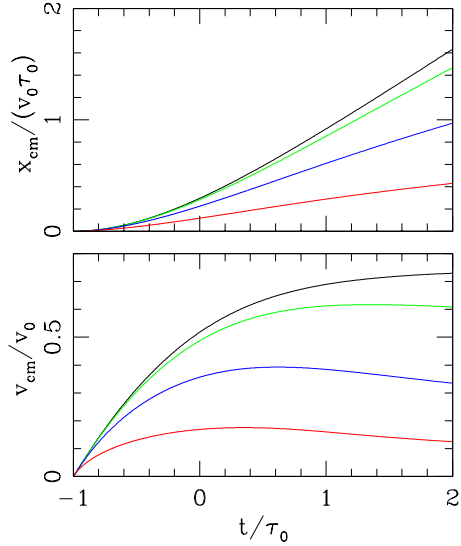}
\caption{Position $x_{cm}$ (top frame) and velocity $v_{cm}$ (bottom frame)
  of the head of the plasmon as a function of time for models with
  $\sigma=0$ (top curves), 0.1, 1.0 and 10 (bottom curves).}
\label{fig2}
\end{figure}

\section{The $\sigma> 0$ case}

For $\sigma>0$, equation (\ref{xcm3}) can be inverted to obtain:
\begin{equation}
  \frac{x_{cm}}{v_0\tau_0}=\frac{1}{2\sigma}
  \left[-f\left(\frac{\tau}{\tau_0}\right)+\sqrt{f^2\left(\frac{\tau}{\tau_0}\right)
      +4\sigma g\left(\frac{\tau}{\tau_0}\right)}\right]\,.
  \label{xcm6}
\end{equation}
The centre of mass positions and velocities as a function of $t$ obtained
for differenent $\sigma$ values are shown in Figure 2.

For the $\sigma=0.1$ case (see Figure 2), $x_{cm}$ and $v_{cm}$ initially
follow the $\sigma=0$ solution (see equation \ref{xcm5}), and start deviating
for $t>0$, when the plasmon head begins to brake in an appreciable way. The
$\sigma=1$ and 10 solutions show substantial braking for all $t$.

The $\sigma>0$ solutions show plasmon heads that first accelerate, then reach
a maximum velocity, and subsequently brake for increasing times $t$. For
$\sigma\ll 1$, the plasmon head first reaches a velocity similar to the
asymptotic velocity $v_a$ of the ``free plasmon'' (see equation \ref{va}) and
then slowly slow down for increasing times. For $\sigma>1$, the velocity of the
plasmon head does not reach values $\sim v_a$.

We should note that in the $\sigma>0$ solutions, all of the mass ejected in the
pulse eventually ends up in the plasmon head.

\section{Position-velocity diagrams}

Evidently, the ``head/tail plasmon'' model is attractive for trying to explain
fast moving clumps which have a tail of decreasing velocity emission towards
the outflow source. When observed with spatially resolved spectroscopy
or with interferometric millimeter observations these clumps
show position-velocity (PV) diagrams
with a high velocity, compact emission at a given position, and a ramp
of emission with increasing radial velocities from the source out to the
clump.

In Figure 3 we show the positions
and velocities of the head at different times, and the velocity
of the material in the ``tail'' of the plasmon. This
velocity is directly obtained from the free-streaming relation:
\begin{equation}
  u(x,t)=u_0(\tau)=\frac{x}{t-\tau}\,,
  \label{uxt5}
\end{equation}
where $u_0(\tau)$ is given by equation (\ref{u0q}). This can be easily
done in a parametric way by varying $\tau$ (at a fixed evolutionary
time $t$), using the first equality to calculate the velocity $u(x,t)$
and then the second equality for obtaining the corresponding position
$x$ along the tail of the plasmon.

Figure 3 shows
the PV diagrams obtained for different values of $t$ for four
models with $\sigma=0$, 0.1, 1 and 2. {The
$\sigma=0$ model has a PV diagram that becomes more extended
along the outflow axis with time, with a plasmon head that
shows a decreasing acceleration for increasing times.}

The models with higher $\sigma$ values have PV diagrams that have
lower peak velocities as a function of $t$.
Regardless of the value of $\sigma$, the ``head/tail'' plasmons
develop an almost linear velocity vs. distance ``Hubble law''
velocity profile at evolutionary times $t\gg\tau_0$. This result
follows from equation (\ref{uxt5}), which in the
$t\gg\tau\sim \tau_0$ limit gives $u(x,t)\approx x/t$ (i.e.,
at a given time $t$ we have  a ``Hubble law'' of slope $1/t$).

\begin{figure}
\centering
\includegraphics[width=6cm]{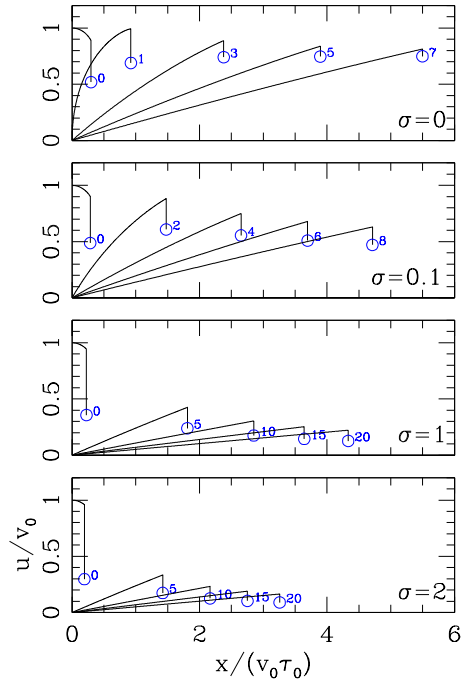}
\caption{Velocity along the outflow axis vs. distance from the outflow
  source at different evolution times. The plots are labeled with the
  value of $\sigma$ of the model (from $\sigma=0$ at the top to $\sigma=2$
  on the bottom). The $\sigma=0$ frame (top graph) shows
  the velocity along the tail as a function of $x$ for times $t/\tau_0=0$
  (shortest curve), 1, 3, 5 and 7 (spatially more extended curve).
  The $\sigma=0.1$ frame shows the velocity vs. position at times
  $t/\tau_0=0$, 2, 4, 6 and 8. The $\sigma=1$ and 2
  frames (two bottom graphs) show the velocity vs. position at times
  $t/\tau_0=0$, 5, 10, 15 and 20. The open circles located at the end of each
  curve show the position and velocity of the head of the plasmon.}
\label{fig3}
\end{figure}

\section{Numerical simulation}

We have computed an axisymmetric gasdynamic simulation
of a ``head/tail plasmon'' with parameters for a high-velocity
clump in a PN (see, e.g., Alcolea et al. 2001)
using the {\sc Walicxe-2D} code (Esquivel et
al. 2009). We use a setup with an adaptive mesh
with 5 refinement levels giving a maximum resolution of
14.64~AU in a computational domain of
$15000 \times 3750$~AU. We used a reflective boundary condition
on the symmetry axis and free outflow for all of the other boundaries.

The ejection velocity pulse is imposed at $x=0$, with a radius $r_j=10^{16}$~cm,
a time half-width $\tau_0=50$~yr and a peak velocity $v_0=200$~km~s$^{-1}$
(see equation \ref{u0q}).
The total mass of the pulse is $M_p=10^{-4}$~M$_\odot$. For calculating
the ejection density, we impose a constant mass loss rate
per unit area ${\dot m}=M_p/(2\pi r_j^2\tau_0)=2.0\times
10^{-13}$~g~cm$^{-2}$~s$^{-1}$, and calculate the density as:
\begin{equation}
  \rho_0(\tau)=\frac{\dot m}{\max\, [u_0(\tau),v_{min}]}\,,
  \label{rho000}
\end{equation}
with $v_{min}=1$~km~s$^{-1}$ ($v_{min}$ is introduced in order to avoid
the divergence of the density for $u_0\to 0$).
Initially, the computational domain is filled
with a uniform environment of numerical density
$n_a=1963.3$ cm$^{-3}$, which, combined with the properties of the
pulse, gives $\sigma=0.1$ (see equation \ref{acm1}). Both the environment
and the ejected material have an initial temperature of $10^4$~K, and
have singly ionized H.

\begin{figure}
\centering
\includegraphics[width=0.7\columnwidth]{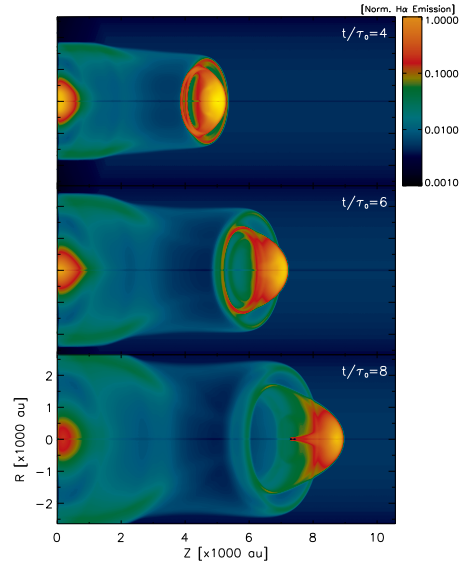}
\caption{H$\alpha$ maps obtained from the numerical simulation
  for evolutionary times $t/\tau_0$=4, 6 and 8 (top, middle and bottom panels, respectively),
  assuming a $\phi=30^\circ$ angle between the outflow axis and the plane of the sky.
  The maps are normalized to the peak emission of the top frame, and are shown with
  the logarithmic colour scale given by the bar to the right of the top frame.}
\label{fig4}
\end{figure}

In the simulation, a minimum temperature of $10^4$~K is imposed
in all cells at all times (also assuming that H is always fully ionized),
and the parametrized cooling function of Biro \& Raga (1994) is used for
$T>10^4$~K. This setup is meant to approximate the behaviour of the gas within
a photoionized region. {Throughout our simulation, the bow shock has a shock
velocity $\sim 100$~km~s$^{-1}$, which together with the pre-shock ambient
density ($n_a\approx 2000$~cm$^{-3}$, see above) gives a cooling distance
$d_c\sim 1$~AU to $10^4$~K (from the plane-parallel shock models of Hartigan et al. 1987),
which is unresolved in our simulation. The slower ``jet shock'' develops velocities
as low as $\sim 20$~km~s$^{-1}$, and does not have substantial cooling in this regime.}

From this simulation, we have calculated predicted
H$\alpha$ maps and PV diagrams. These are obtained by computing the H$\alpha$
emission coefficient (using the interpolation of Aller 1994), and
integrating it through lines of sight.

Figure \ref{fig4} shows the H$\alpha$ emission maps obtained
for evolutionary times $t/\tau_0$=4, 6 and 8 (upper, middle and
bottom panels, respectively), assuming a $\phi=30^{\circ}$ angle
between the outflow axis and the plane of the sky. From this figure
we see that the H$\alpha$ emission has two components: the plasmon
head and the tail. This latter component is brightest close to the
outflow source. {The bow shock at the head of the plasmon is rather broad,
which is a result of the fact that the Mach number of the flow is not so high
(going down to $\sim 10$ towards the end of the simulation).}

\begin{figure}
\centering
\includegraphics[width=0.8\columnwidth]{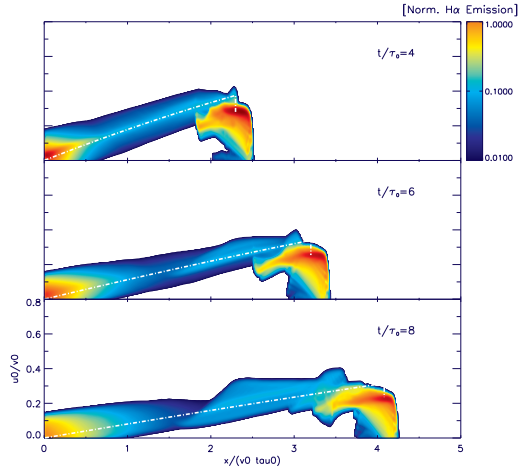}
\caption{Position-velocity (PV) diagrams obtained from the simulation
  for evolutionary times $t/\tau_0$=4, 6 and 8 (top, middle and bottom panels, respectively),
  assuming a $\phi=30^{\circ}$ angle between the outflow axis and the plane of the
  sky. The PV diagrams
  are normalized to the peak emission of the top frame, and are shown with
  the logarithmic colour scale given by the bar to the right of the top frame.
  The (appropriately projected) velocity vs. position obtained from the
  analytical solution (for the corresponding evolutionary times) is shown with
  the dashed white curves.}
\label{fig5}
\end{figure}

We also calculate the PV diagrams for evolutionary times
$t/\tau_0$=4, 6 and 8, and a $\phi=30^{\circ}$ angle between the
outflow axis and the plane of the sky (see Figure \ref{fig5}).
For the PV diagrams, we have assumed that we have a spectrograph slit
with a full width of 100~AU, centred on the outflow axis. The
resulting PV diagrams show a clear ``Hubble law'' ramp of increasing
radial velocities vs. distance from the source, ending in a broad
emission line region corresponding to the head of the plasmon.

In Figure \ref{fig5} we also plot
the (appropriately projected) velocity
vs. position obtained from the analytical model
(see section 5 and Figure \ref{fig3}). {The "Hubble law" feature of
the tail agrees very well with the results obtained from the numerical
simulation. Also, the analytic position of the plasmon head falls in
the middle of the spatially quite extended emission predicted from the
numerical simulation (this spatial extent being partly the result of the
projection of the wide bow shock onto the plane of the sky).}

\section{Conclusions}

We present a model for a hypersonic ``single pulse jet'', produced by a collimated
outflow event with an ejection velocity history with a single peak, and
wings of decreasing velocity (at earlier and later times). An arbitrary
form for a simultaneous ejection density variability is also possible.

Such an ejection results in the formation of a ``head'' associated with
a working surface travelling through the surrounding environment, and a
``tail'' of slower material (formed by the decaying velocity tail of
the outflow event) which rapidly develops a linear, ``Hubble law''
kinematical signature. We call this flow configuration a ``head/tail plasmon''.

We study the simple case of a parabolic ejection velocity pulse (which
could be viewed as a second order Taylor series of the peak of an arbitrary
ejection pulse), with a time-independent mass loss rate (i.e.,
the ejection density is proportional to the inverse of the ejection velocity).
With a ``centre of mass formalism'', we obtain the motion of the head
of the ``head/tail plasmon'' (see section 2).

In the limit of a very low density environment (see section 3)
the head of the plasmon reaches a constant velocity, and the material in the
tail at all times retains a substantial fraction (asymptotically
approaching 1/4) of the total mass of the ejection event. For denser
environments (see section 4), the plasmon slows dowm, and the
head ends up incorporating most of the mass of the ejection pulse.
For all flow paramenters, the predicted PV diagrams rapidly develop
a ``Hubble law'' kinematical signature (see section 5).

Finally, we compute an axisymmetric gasdynamic simulation with parameters
appropriate for a high velocity clump in a PN (see section 6).
We compute H$\alpha$ emission maps and PV diagrams showing the observational
characteristics this flow.
The predicted PV diagrams obtained from the simulation agree very well with
the analytic model (see Figure 5).

This paper represents a first exploration of a different kind of jet or plasmon
flow. A detailed application
of this model to different objects will be necessary to show what improvements
are found with respect to previous models, such as the ones of Dennis et al.
(2008) for knots in PNe, or the ones of Rivera et al. (2019a, b)
for the Orion BN-KL fingers.

\section*{Acknowledgments}
We acknowledge support from the PAPIIT (UNAM) project
IG100218/BG100218. ACR acknowledges support from a DGAPA-UNAM posdoctoral
fellowship. We thank an anonymous referee for helpful comments.

\noindent Data availability: The lead author may be
contacted for access to the results of the simulations.

\bsp

\label{lastpage}

\end{document}